\documentstyle[aps,prl,multicol,epsf]{revtex}

\begin{document}

\title{A Model for Bone Strength and Osteoporotic Fractures}

\author{Gemunu H. Gunaratne,$^{1,2}$ Chamith S. Rajapaksa,$^{1}$ Kevin E. Bassler,$^{1}$
Kishore K. Mohanty,$^{3}$ and Sunil J. Wimalawansa$^{4}$}

\address{$^{1}$ Department of Physics,
                University of Houston,
                Houston, TX 77204}
\address{$^{2}$ The Institute of Fundamental Studies,
                Kandy 20000, Sri Lanka}
\address{$^{3}$ Department of Chemical Engineering,
                University of Houston,
                Houston, TX 77204}
\address{$^{4}$ Department of Internal Medicine,
                University of Texas Medical Branch,
                Galveston, TX 77555}
\maketitle
\nobreak

\begin{abstract}
Inner porous regions play a  critical role in the load bearing capability of large bones.
We show that an extension of disordered elastic networks [Chung et. al., Phys. Rev. B,
{\bf 54}, 15094 (1996)] exhibits analogs of several known mechanical features of bone. 
The ``stress-backbones" and histograms of stress distributions for healthy 
and weak networks are shown to be qualitatively different. A hereto untested relationship
between bone density and bone strength is presented.
\end{abstract}

\nobreak
\begin{multicols}{2}

Osteoporosis is a multi-faceted metabolic disease that reduces bone strength
and leads to a significant number of fractures occurring in older
adults~\cite{marAjoh,fung}. Unfortunately, therapeutic agents available
for prevention and treatment of osteoporosis often induce
adverse effects in patients~\cite{wein}.
Thus, non-invasive diagnostic tools to determine the need for therapeutic intervention 
are essential for effective management of osteoporosis. 
Simple models can form a useful complement to traditional
studies of osteoporosis. In this 
Letter we introduce such a model, and describe some of its features.

Large bones such as thigh bones and vertebrae consist of an outer cylindrical shaft 
(cortex) and an inner porous region (trabecular architecture)~\cite{fung}. The cortex
is made of compact bone and has a thickness of several millimeters. The structure of the
trabecular architecture (TA) is that of a disordered cubic network of 
``trabeculae" whose axial and cross sectional 
dimensions are of the order 1mm and 0.1mm respectively~\cite{fung} (see Figure 1).

Routine activities (e.g., climbing stairs) inflict
micro-damage on bone. Material in the neighborhood of 
these fractures is resorbed by a class of cells known as ``osteoclasts." Their
presence attracts a second group of cells, ``osteoblasts," 
which help regenerate lost bone~\cite{fung}. This
sequence of events, referred to as {\it bone remodeling}, reduces the accumulation of 
micro-damage shallower than about 0.1mm~\cite{wein,norAyen}. The full restoration of 
bone strength typically takes a period of 2-3 months.

In the cortex, deeper fractures, created during occasional trauma, are not repaired through
remodeling. The resulting micro-fracture accumulation leads to lower bone quality 
and fracture toughness~\cite{norAyen,finAmar}, reducing the load bearing capacity of
the cortex with aging. 

Since the thickness of trabecalae is $\sim$0.1 mm, fractures that do not sever 
them can be remodeled.  Clinical studies also indicate that perforated 
trabeculae are seldom regenerated~\cite{parAmat} and that 
cross sections of those surviving changes little~\cite{legAcha}. 
Thus, even though the connectivity of the TA reduces with aging, the surviving
trabeculae can (mostly) be expected to retain their quality. Indirect evidence for this
conjecture is provided by mechanical studies which have shown that a TA
from a given skeletal location fractures at a {\it fixed} level of
strain {\it independent of the age of the bone}~\cite{hogAruh}. (In 
contrast, the breaking stresses of such samples reduce with age, though much less
than the corresponding degradation of a cortex.) Due to efficacy of its remodeling, the 
TA becomes the principal load carrier in bones of older adults. 

Reductions in estrogen and testosterones lead to increased bone turnover and
an imbalance in remodeling.  The accompanying weakening of a TA increases the
susceptibility of bone to fracture~\cite{potAben}. The management of osteoporosis 
can be greatly aided by the availability of characteristics that can identify this weakening. 

\begin{figure}
\narrowtext
\epsfxsize=2.00truein
\hskip 0.6truein
\epsffile{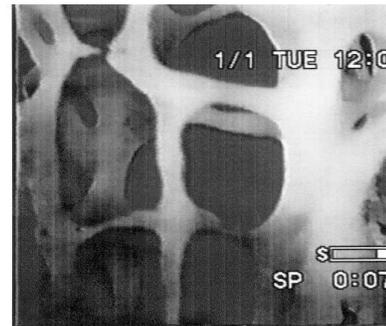}
\caption{A cross section of the TA from a 70 year old female. Observe that
the trabeculae form a disordered cubic network. The network loses connectivity with aging.} 
\label{real_bone} 
\end{figure}

Bone Mineral Density (BMD), or the effective bone density is the principal clinical measurement 
used as a surrogate of  bone strength~\cite{cann}. It is estimated using
x-ray imaging via an evaluation of the opacity of bone. Large clinical studies
have shown an exponential increase in fracture incidence with the loss
of BMD~\cite{reck}. Among other features known to be relevant for bone strength 
is the architecture of the porous bone. Structural properties of a TA such as the 
average width of the trabeculae, mean 
trabecular spacing, connectivity of the network and
its fractal dimension have been proposed as additional surrogates of
the  ultimate (or breaking) strength of bone~\cite{legAcha,potAben,majAnew}.
Alternative approaches to estimate bone strength involve the use of 
finite element computations~\cite{ladAkin} and cellular models~\cite{vajAkra}.
All previous analyses have been carried out on TAs from bone samples. This makes it 
nearly impossible to isolate effects of individual factors on bone decay.

We introduce a simple mechanical model of a TA to complement these analyses. 
Since model parameters can be varied independently, we expect
the task of  elucidating the  essential differences
between healthy and osteoporotic bone to be simplified. 
Possible diagnostic tools for osteoporosis can then be identified 
using their ability to quantify these differences.

Images (e.g., Figure 1) suggest that disordered elastic networks~\cite{chuAroo} 
may be used to model TAs.
Our initial studies are conducted on a square network of linear springs. 
An unstressed configuration is generated by displacing the vertices of the 
square grid (of side $D$)  randomly by an amount less
than $D \Delta$. Externally imposed deviations increase the potential energy of the network 
via a combination of elastic ($\frac{1}{2} k (\Delta l)^2$) and bond 
bending ($\frac{1}{2} \kappa (\Delta \theta)^2$) contributions. Here $\Delta l$ and 
$\Delta \theta$ are changes in the length of a spring and the bond angle between 
adjacent springs respectively~\cite{chuAroo}. The differences between trabeculae 
are modeled by assigning random values for $k$ and $\kappa$;
specifically $k \in [k_0(1-\eta_e), k_0(1+\eta_e) ]$ and 
$\kappa \in [\kappa_0 (1-\eta_b), \kappa_0(1+\eta_b) ]$, where 
$\eta_e$ and $\eta_b$ are predetermined. Since the network is 
constrained to lie on a plane, torsional effects are ignored. 
The following conditions are included to model known features of bone.

\begin{itemize}
\item As discussed above, the fracture criterion for trabeculae should be based on the
level of strain~\cite{hogAruh}. Following mechanical studies of fracture, we assume that 
fracture strains $\gamma$ are distributed on 
a (two-parameter) Weibull distribution~\cite{harAphe,leaAdux}, with cumulative probability  
\begin{equation}
C(\gamma) = 1 - exp \Bigl[-(\gamma / \gamma_e)^m \Bigr].
\end{equation}
An elastic element which is strained beyond its fracture strain is removed from the network along with
all bond-bending forces it contributes to. The bonds are assumed to fracture when changed 
beyond an amount distributed on a second Weibull distribution (with parameters
$\gamma_b$ and $m$); the bond and its side 
with the smaller spring constant are removed from the network.

\item Osteoporosis is modeled by a random removal of springs from the network. 
The probability of removal, $\nu$, is used to quantify the ``level of osteoporosis;" 
links eliminated at a given value of $\nu$ are not re-introduced later. The surviving
springs are assumed to retain their strength.

\item Clinical studies show a dramatic increase of bone strength (disproportionate to the 
increase of BMD) following therapeutic intervention~\cite{reck}. It is
explained by assuming that bone remodeling is preferentially carried out 
in locations of high stress (Wolff's Law~\cite{wolf}). In the model, 
elasticity of the spring under maximum {\it strain} is increased by a small
fraction $p$; those strained by a
fraction $f> f_0$ of the maximum are strengthened by $p \bigl(\frac{f-f_0}{1-f_0}\bigr)^{\beta}$.
This algorithm is implemented over repeated application of a given external strain.
\end{itemize}

\begin{figure}
\narrowtext
\epsfxsize=2.50truein
\hskip 0.3truein
\epsffile{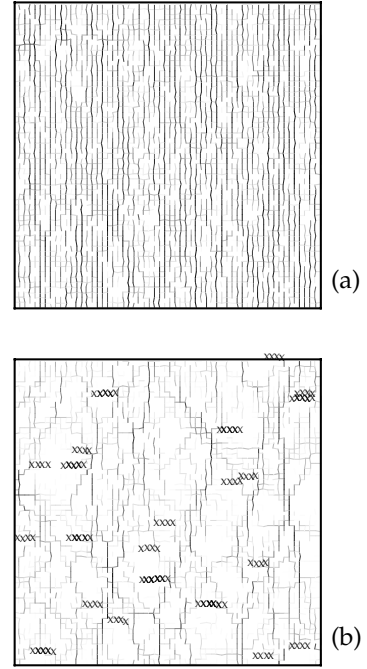}
\vskip 0.20in
\caption{Stress distributions on $60\times 60$ networks
with $D=10.0$, $\Delta=0.1$, $k_0=1$, $\kappa_0=5$, $\eta_b=0.5$, $\eta_a =
0.5$, $\gamma_e=0.05$, $\gamma_b=0.1$, and $m=5$. For clarity only the compressed springs are shown, 
and elements under higher stresses are represented by darker
hues. (a) When a network with $\nu=15\%$ is strained by $\zeta=4.0$, 
the stress backbone is evenly distributed. (b) In contrast, the stress backbone when 
$\nu=30\%$ ($\zeta=4.0$) consists of a few coherent pathways. The crosses denote
locations with large fractures in the x-direction.}
\label{backbone}
\end{figure}

Below, we discuss properties of networks subjected to uniform strain $\zeta$, 
reached through a sequence of small, adiabatic increments~\cite{conjgrad}. 
When elastic elements are removed from the network due to fracture, equilibrium is 
recalculated prior to increasing $\zeta$. 
The sides of the network are constrained to move vertically,
as is the case for a TA because it is connected to the cortex. The force $T(\zeta)$
required to sustain a given strain is estimated by adding the vertical 
forces on the upper surface.  Results analogous to those presented are observed 
under four-point bending which is a leading mechanism of 
fracture of long bones~\cite{bouAaug}.

The stress-strain relationship $T(\zeta)$, which is initially linear
becomes nonlinear beyond the ``yield point" and reaches a maximum (ultimate stress) 
$T_{max}$ prior to failure of the network.  The yield point coincides with the first 
fracture of springs.  Indirect evidence has been presented to 
suggest that yielding of a bone coincides with cracking of trabeculae~\cite{keaAguo}.
Failure of a network is accompanied by a crack propagating across the entire 
network.

Figure~\ref{backbone}(a) shows the stress distributions on a ``healthy" network  
strained below yield. Springs experiencing large stresses (forming the
``stress backbone") are distributed evenly throughout the network.
The histogram of stresses for this configuration contains a broad peak, similar to histograms for 
elastic networks~\cite{chaAmac}.  When $\zeta$ is increased, the stress backbone occupies a  
smaller subset following fracture of a group of springs.
Similar changes  have been observed in elastic networks~\cite{sahAarb} and in finite 
element computations on digitized images of bone~\cite{ladAkin}. 

The nature of the stress backbone is very different in weaker networks; i.e., 
those with larger values of $\nu$. As shown in Figure~\ref{backbone}(b), the
stress backbone for such networks consists of a few coherent pathways. 
Beyond $\nu > \nu_0 \approx 20\%$ there is no peak in the stress-histogram. 
We expect these differences to
prove useful in identifying new diagnostic tools for osteoporosis.

In Figure~\ref{backbone}, the X's denote long horizontal fractures; specifically, 
locations where four or more consecutive vertical bonds are absent. 
They prevent the propagation of stress in significant vertical slices, 
even when bond-bending forces are included. When $\nu$ is large, the stress
backbone can be seen to avoid these regions, and consequently contains only a 
small fraction of available bonds.

The last observation can be used to estimate the decay of bone strength with increasing
$\nu$. Since an externally applied stress passes through every
horizontal layer, we need to consider stress propagation on each one dimensional slice.
This ``chain-of-bundles" model~\cite{leaAdux} further assumes that the stress on each edge of a 
fracture containing $k$ sites is enhanced by a factor $(1+\frac{k}{2})$; i.e., the edges share the 
load assigned to bonds in the fracture.

\begin{figure}
\narrowtext
\epsfxsize=2.00truein
\hskip 0.5truein
\epsffile{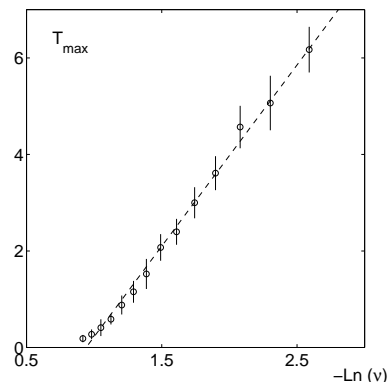}
\vskip 0.02in
\caption{The mean and standard error of the ultimate stress $T_{max}$
for five networks with identical control parameters as a function of $\nu$. The dashed
line is the least squares fit.}
\label{ultstr} 
\end{figure}

Consider an elastic network of size $N\times N$. The probability of forming a fracture of length $k$ 
(with links on either side) is $\nu^k (1-\nu)^2$. Hence the size of the largest 
fracture $k_m$ (which occurs with a probability of $\sim 1$~\cite{kahAbat}) can be estimated by
\begin{equation}
N^2 \nu^{k_m} (1-\nu)^2 \sim 1 .
\label {frac_size}
\end{equation}
For sufficiently large fractures $k_m \sim -2 \ln N / \ln \nu$. In the 
chain-of-bundles model, the stress on each side of this fracture is 
$(1+\frac{k_m}{2})\frac{T}{N}$, where $T$ is
the externally applied stress. When this stress is the typical breaking stress
of an elastic element, the fracture will propagate. Thus 
$T_{max}$ is related to $\nu$ through
\begin{equation}
  T_{max} \sim \frac{1}{k_m} \sim -\ln \nu.
\label {decay2d}
\end{equation}
Figure~\ref{ultstr} shows that the decay of $T_{max}$ in model networks
is consistent with this expression.

A similar analysis of ``penny-shaped" fractures in three dimensional networks 
gives 
\begin{equation}
T_{max} \sim \sqrt{-\ln \nu}.
\label {decay3d}
\end{equation}
Results from mechanical studies of bone samples (with an assumed  power law relationship)
show that $T_{max}\sim (BMD)^{-\alpha}$, where
$\alpha \approx 2.6$~\cite{mosAebb}. There is no theoretical basis for 
this power law, and Eqn. (\ref{decay3d}) also provides a satisfactory 
fit to the data. More sensitive experiments need to be conducted to
discriminate between these possibilities.

\begin{figure}
\narrowtext
\epsfxsize=2.00truein
\hskip 0.5truein
\epsffile{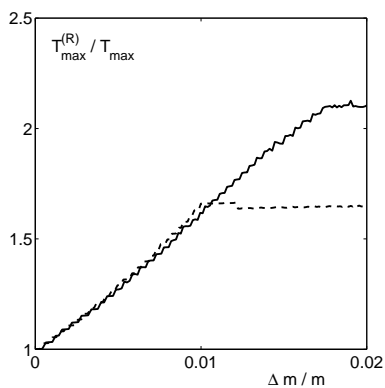}
\vskip 0.02in
\caption{The fractional enhancement of $T_{max}$ as a function of the percentage
increase of BMD under therapeutic regeneration
(with $\beta=2$, $f_0=0.8$) for two  networks with 
$\nu=10\%$ (dashed line) and $\nu=20\%$ (solid line).}
\label{regen} 
\end{figure}

We finally discuss ``therapeutic regeneration" (introduced above) of a 
network. There has not been quantitative studies on this issue, but it has been 
suggested that therapeutic regeneration will be more effective in increasing the
strength of {\it healthy} bone~\cite{mosAebb}. This suggestion is based on an assumed
absence of hysteresis with changes in BMD. 
The validity of this proposal can be tested in the model system~\cite{foot5}. 
Figure~\ref{regen} shows the fractional increases of $T_{max}$ experienced by 
two networks that are formed by degrading the same lattice by $\nu=10\%$ and $\nu=20\%$. 
As seen in clinical studies, there is a dramatic enhancement of
bone strength ($\sim$ factor 2), disproportionate to the increase of BMD ($\sim 2\%$).
For the example shown, the weaker network is seen to strengthen by a larger factor under 
increase of BMD. There are large fluctuations of the enhancement between distinct networks, 
and in occasional examples the growth of the stronger network is larger.

In this Letter, we have argued that an extension of elastic networks 
can be used to model the inner porous regions of 
bone. Since the latter is the principal load carrier in bones of older adults,
the model can prove useful in identifying characteristics of osteoporotic bone. 
Eqn. (\ref{decay3d}) is compatible
with published data, but more sensitive tests are needed to discriminate it
from a possible power law decay. Examples from the model provide evidence that
bone regeneration under Wolff's law is not compatible with the absence of
hysteresis in bone strength with bone density. In fact, generally, regeneration
is more efficient in strengthening networks that are weaker.

We propose to use the model identify new diagnostic tools for osteoporosis,
using characteristics that can differentiate between stress
backbones (Figure~\ref{backbone}). Preliminary studies 
indicate that the ratio of responses of a network to static and periodic strain is a
suitable surrogate of bone strength~\cite{guna}.

The authors would like to thank M. P. Marder for discussions on fracture and 
S. R. Nagel for suggesting the possible use of response measurements to 
quantify variations in stress backbones. We would also like to thank the referees 
of earlier drafts of the manuscript for their comments.
This research is partially funded by the Office of Naval Research
(GHG), National Science Foundation (GHG and KEB), Texas Higher Education
Coordinate Board (GHG, KEB, KKM and SJW), Department of Energy (KKM), and
National Institutes of Health (SJW).

\end{multicols}

\begin{thebibliography}{99}
\bibitem{foot5} We estimate the increase of BMD by the enhancement of the average 
spring constant of the network under regeneration.  In a lattice consisting of uniform rods, 
the mean spring constant is proportional to the mass.
\end{thebibliography}
\end{document}